\documentclass[11pt,a4paper]{article} 
\usepackage{jheppub}
\usepackage{graphicx}%
\usepackage{multirow}%
\usepackage{amsmath,amssymb,amsfonts}%
\usepackage{amsthm}%
\usepackage{mathrsfs}%
\usepackage[title]{appendix}%
\usepackage{xcolor}%
\usepackage{textcomp}%
\usepackage{manyfoot}%
\usepackage{booktabs}%
\usepackage{algorithm}%
\usepackage{algorithmicx}%
\usepackage{algpseudocode}%
\usepackage{listings}%
\usepackage[utf8x]{inputenc}     
\usepackage{enumerate}
\usepackage{epsfig} 
\usepackage{float} 
\usepackage[caption = false]{subfig}
\usepackage{wasysym} 
\usepackage{mathrsfs} 
\usepackage{amsfonts} 
\usepackage{amsbsy} 
\usepackage{amscd} 
\usepackage{pstricks} 
\usepackage{multirow} 
\usepackage{tikz}
\usepackage{color}
\usepackage{slashed}
\usepackage{array}
\usepackage{tablefootnote}
\usepackage{amsmath}

\usetikzlibrary{arrows,positioning,shapes.geometric} 
\usepackage[compat=1.1.0]{tikz-feynman}          
\usepackage[font=small,labelfont=bf]{caption}
\usepackage{slashed}
\usepackage{multirow}
\usepackage{tabularx}
\usepackage{soul}  
\usepackage{listings} 
\usepackage{color}

\usepackage{amsthm}

\begin{document}
\title{Foundations of automatic feature extraction at LHC--point clouds and graphs
}

\author[a]{Akanksha Bhardwaj,}
\author[b]{Partha Konar,}
\author[c]{and Vishal~S.~Ngairangbam}
\affiliation[a]{Department of Physics, Oklahoma State University, Stillwater, OK, 74078, USA}

\affiliation[b]{Theoretical Physics Division, Physical Research Laboratory,\\ Shree Pannalal Patel Marg, Ahmedabad - 380009, Gujarat, India}
\affiliation[c]{Institute for Particle Physics Phenomenology, Department of Physics,\\ Durham University, Durham DH1 3LE, United Kingdom}

\emailAdd{akanksha.bhardwaj@okstate.edu}
\emailAdd{konar@prl.res.in}
\emailAdd{vishal.s.ngairangbam@durham.ac.uk}

\abstract{Deep learning algorithms will play a key role in the upcoming runs of the Large Hadron Collider (LHC), helping bolster various fronts ranging from fast and accurate detector simulations to physics analysis probing possible deviations from the Standard Model. The game-changing feature of these new algorithms is the ability to extract relevant information from high-dimensional input spaces, often regarded as ``replacing the expert" in designing physics-intuitive variables. While this may seem true at first glance, it is far from reality. Existing research shows that physics-inspired feature extractors have many advantages beyond improving the qualitative understanding of the extracted features. In this review, we systematically explore automatic feature extraction from a phenomenological viewpoint and the motivation for physics-inspired architectures. We also discuss how prior knowledge from physics results in the naturalness of the point cloud representation and discuss graph-based applications to LHC phenomenology.

}
\preprint{IPPP/24/22}

\allowdisplaybreaks

\maketitle
\section{Introduction}

Modern machine learning (ML) techniques are quickly becoming ubiquitous in the natural sciences due to their excellent data processing capabilities and ability to find excellent interpolations from high-dimensional data. The situation is not too different in various branches of physics (see ref.\cite{RevModPhys.91.045002}  for a recent review), where they are being employed to enhance traditional ways of studying physical systems ranging from finding the ground state of many-body quantum systems to predicting cosmological parameters. The situation is similar in collider phenomenology. Particularly relevant for the huge data that the Large Hadron Collider (LHC) will generate in the high-luminosity runs~\cite{Elmsheuser:2020omj,HEPSoftwareFoundation:2020daq,Held:2024gwj,Motta:2024wyo}, the community is currently striving to work out the intricacies of these efficient data handling algorithms ranging from their applications to triggers~\cite{Nguyen:2018ugw,Duarte:2018ite,Butter:2022lkf,Bortolato:2024uqg}: the stage which decides which events to store and which to discard, the study of various detector modules~\cite{Paganini:2017hrr,Farrell:2018cjr,Qasim:2019otl,ATL-SOFT-PUB-2020-006,Buhmann:2020pmy,Biscarat:2021dlj,calochallenge,Adelmann:2022ozp,Liu:2023siw}, phenomenological applications~\cite{Baldi:2014kfa,Dery:2017fap,Metodiev:2017vrx,Kasieczka:2021xcg,ATLAS:2020iwa,Aarrestad:2021oeb,Hallin:2022eoq,ATLAS:2023ixc,Ngairangbam:2023cps,CMS:2024lwn}: which connects the recorded data to update our current understanding of the physical universe, and foundation models~\cite{Finke:2023veq,Butter:2023fov,Vigl:2024lat,Heinrich:2024sbg,Birk_2024,Harris:2024sra}  capable of generalising across several applications.

The defining characteristic of current state-of-the-art ML algorithms is their ability to extract features from a high-dimensional input representation, which has undergone little to no processing (in terms of dimensionality reduction to a few variables through domain knowledge). These so-called deep-learning algorithms often need not rely on hand-engineered variables and can (with appropriate architecture design) outperform shallow methods relying on human-designed variables. The downside of such algorithms is the loss of understanding, in the traditional sense, of what the network is exploiting to outperform the human-engineered variables. Even though the increase, in principle, can be attributed to a higher degree of optimisation via numerical methods coupled with the neural network's inherent power, the very high-dimensional parameter space prohibits an analytic understanding of the optimisation process.

\begin{figure}
	\centering
	\includegraphics[scale=0.4]{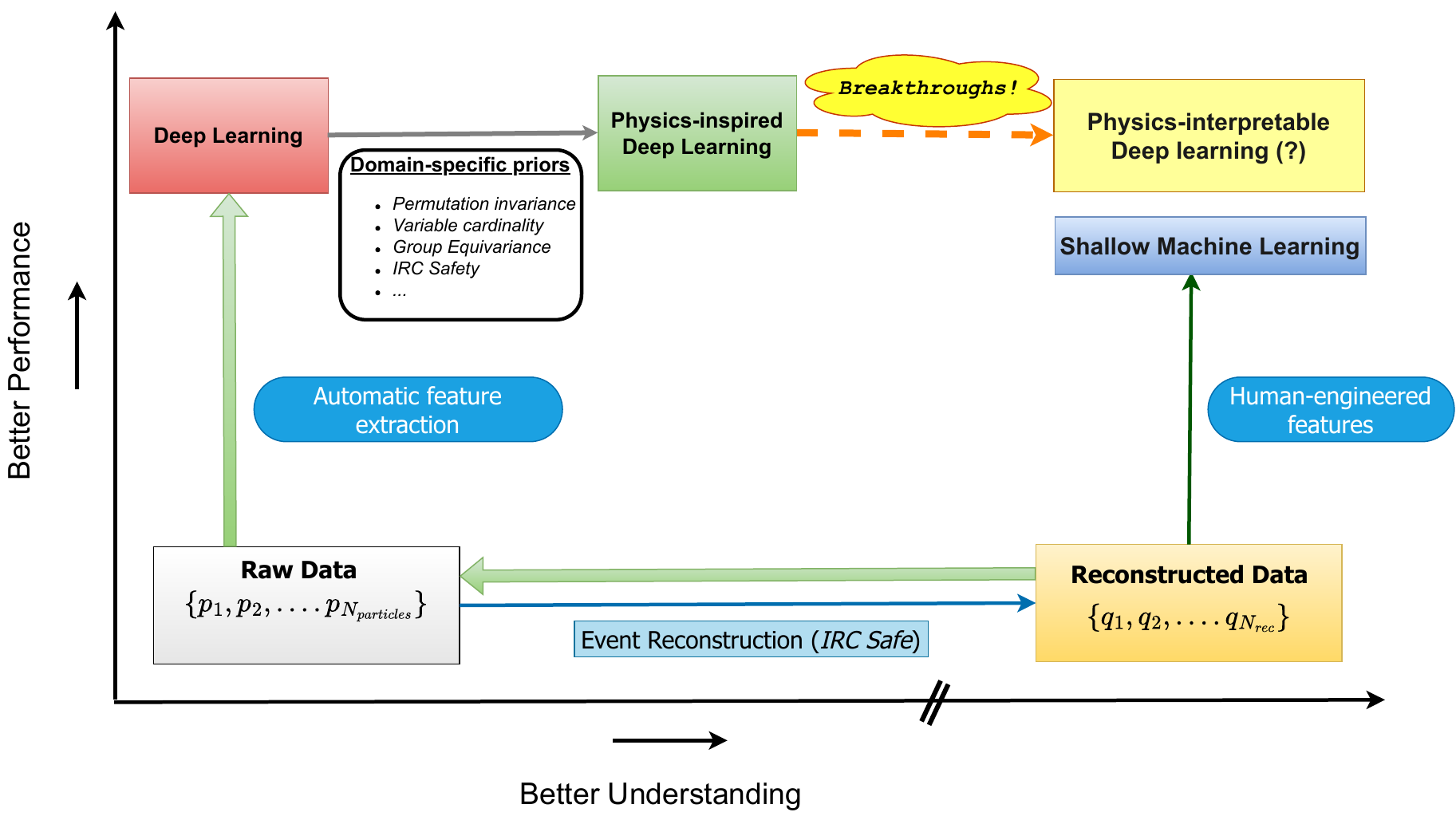}
	\caption{A schematic diagram on the plane of classifier performance and understanding of deep learning algorithms and its comparison to shallow machine learning methods.}
	\label{fig:interpret_deep} 
\end{figure}

Taking the example of signal vs background classification as a proxy for performance, we can cast the interplay of performance and physics-understanding in a two-dimensional plane as shown in figure~\ref{fig:interpret_deep}. In the bottom left, with virtually no separation and understanding reside the raw detector data\footnote{Note that what we consider raw data here is still highly processed when one considers the intricacies of experimental measurements like track finding, pileup subtraction etc.} used as inputs to reconstruction algorithms.  This process which include defining infrared and collinear safe jets, and other objects like leptons, photons, etc., based on isolation criteria on the measurements of the various components of the detector, give us a set of reconstructed objects on the bottom right. On the reconstructed data, one generally applies loose selection cuts to enrich the signal and then apply shallow machine learning after constructing observables based on physics insights of the specific signature in question. In deep-learning based analyses, one goes back to the raw data after pre-selection and uses the high dimensional data as inputs to a suitable model. These models generally reside higher on the performance axis compared to shallow methods but fare poorly on the understanding axis. With physics-inspired deep-learning methods, there is a small increase on the understanding axis with still comparable performance to out-of-the-box methods. However, this is still a long way from the traditional understanding of physics-based observable design (evident from the discontinuous scale on the understanding axis). It would be considered a breakthrough to reach an equal level of understanding to traditional approaches without potentially losing out on the performance gain. Although, what qualifies as equal, will of course be decided by the community at large, a pragmatic criterion could be the understanding of their systematic uncertanties~\cite{NIPS2017_48ab2f9b,Kasieczka:2021xcg,Araz:2021wqm,Ghosh:2021roe,Chen:2022pzc,Golutvin:2023fle}, especially those with theoretical origins~\cite{Ghosh:2022lrf,Ghosh:2021hrh}. 

We systematically review the nature of automatic feature extraction for phenomenology at the Large Hadron Collider, taking a more expositionary approach to what is already known in mainstream machine learning literature and their relations/reinterpretations specific to LHC phenomenology. One major theme of current research in this context, is the suitability of the point cloud representation in analysing high-energy collision events. In the point cloud representation, events are described as an unordered set of particles, characterised by their measured properties like four-momentum, charge, etc., whose number is not constant on an event-by-event basis. This is due to the non-conservation of particle multiplicity in relativistic quantum mechanics and our innate interest in how the measured particles relate to the interactions at subnuclear length scales. As the details of these interactions reside in the correlations between the measured particles, relational structures (like a graph) between the set's elements are often employed to expose these correlations. While we touch upon graph-based representations and overview some applications, our coverage is essentially incomplete as our aim is to elucidate the interplay between the expressive power and generalisation capabilities of deep neural networks in the context of high-energy physics. For a more detailed literature review, we refer interested readers to refs.~\cite{Larkoski:2017jix,doi:10.1142/9789811234033_0007,Duarte:2020ngm,Shlomi:2020gdn,Thais:2022iok,Belis:2023mqs,Workman:2022ynf,DeZoort2023,Hashemi:2023ruu,Hashemi:2023rgo} and the living review~\cite{Feickert:2021ajf}.

The rest of the review is organised as follows. In Section~\ref{sec:uat}, we present a qualitative discussion of the expressivity of fully-connected feed-forward neural networks, and its interplay between generalisation performance, arguing the need to curtail the theoretical expressivity to have good generalisation capabilities for specific purposes. We discuss these prior assumptions in the context of LHC phenomenology in Section~\ref{sec:hep_priors} which motivates the set-based point cloud representation.    Thereby, we present an introduction to graph-based architectures acting on the point cloud representation in Section~\ref{sec:gnn}.  In Section \ref{sec:overview}, we overview some applications relevant to LHC phenomenology. We summarise in Section~\ref{sec:summ}. 

\section{Universal Approximation: theory and practice}
\label{sec:uat} 
  Artificial Neural Networks (ANNs) are non-parametric models which learn an underlying target function based on observed data. By non-parametric, we mean that the free parameters do not have any inherent (physical) meaning and the analysis relies on approximating the target function as accurately as possible.  The models are structured to capture a large set of possible functions.  ANNs owe their origin to mathematical models of the biological neuron~\cite{McCulloch1943,rosenblatt_1958}. As we will be discussing point cloud architectures which utilise dense feed-forward networks as building blocks, we will use the term Multilayer Layer Perceptron (MLP) to refer to such networks to avoid confusion with the composite architectures built out of such units.

  This section presents a brief account of the mathematical structure of MLPs, their expressive power captured in terms of Universal Approximation Theorems (UAT). These UATs, generally phrased as existence theorems, deal with the ability to approximate any continuous function on a compact domain up to any arbitrary precision. Such generality at times impede a practical optimisation especially when the input dimensionality is very large. The expressivity is meticulously reduced based on the input representation and the training objective by biasing the network architecture to some smaller sets of functions.  This preconditions are  colloquially referred to as  ``\emph{inductive biases}", 
  closely related to the paradigm of inductive learning. In such scenarios, we train a model on a specific training dataset to be able to generalise to samples not present during the training, assuming that the new samples follow the same relationship between the input and the target variable. The expected error on the unseen samples is called the generalisation error of the model and it is estimated with the help of the validation dataset. Highly expressive models generally result in very good training error with  poor generalisation and frequently overfit the training data. When the architecture is suitable for the particular task at hand, it results in a better generalisation error and hence can reach lower levels of validation loss without overfitting to the training dataset. We end the section with a discussion on the need to curtail the expressivity via such biases.

  \subsection{Structure of Multilayer Perceptrons} 
  A statistical model takes an input vector $\mathbf{x}\in\mathbb{R}^n$ and learns a target function\footnote{Without loss of generality, we will consider a scalar function.} $\tilde{y}=f(\Theta,\mathbf{x})$ by tuning the set of parameters $\Theta$ according to some known pair of tuples $\{(\mathbf{x},y)_i\}$.  Different models have different forms of functional dependence on the parameters and may involve hidden representations as in the case of neural networks. For MLPs, it is an affine map from the input space $\mathcal{X}\ni\mathbf{x}$ to a sequence of spaces $\mathcal{Z}_\alpha$.  Without any additional restrictions, these are simply $\mathbb{R}^{n_\alpha}$, with $n_{\alpha}$ the dimensionality of the $\alpha$-th representation. Denoting the input vector as $\mathbf{x}=\mathbf{z}_0$ and the output as $\tilde{y}=\mathbf{z}_{n+1}$, with $n$ being the number of hidden layers, the mathematical form can be written as 
  \begin{equation}
  	\mathbf{z}_{\alpha}=A(\mathbf{w}_\alpha\,.\,\mathbf{z}_{\alpha-1}+\mathbf{b}_\alpha)\quad,
  	\end{equation}  where $\mathbf{w}_\alpha$ is a $n_{\alpha}\times n_{\alpha-1}$ weight matrix and $\mathbf{b}_\alpha$ is a $n_{\alpha}$ dimensional weight vector. $A$ is an \emph{activation function} applied to each element of the argument vector.  As composition of of linear functions is a linear function, activation functions need to be be non-linear to be able to capture non linearities in data. On the other hand, not every non-linear activation usually leads to the universal approximation property, although the requirements are not very restrictive as we shall see in the following discussions.    
  	
  	The architecture as described above, are sequential, i.e., each layer feeds to the next and so on, to give the final predicted output, or the function $\tilde{y}(\mathbf{x})$ is a composition of several functions. In cases where the quality of the approximation can be quantified with a differentiable metric between $\tilde{y}$ and the true value $y$, one generally uses a gradient descent algorithm to optimise the parameters $\mathbf{w}_\alpha$ and $\mathbf{b}_\alpha$ which are initialised to some random value before the training starts.\footnote{For more details of how the process of initialisation affects the network performance, see for instance refs.~\cite{pmlr-v9-glorot10a,7410480,Narkhede2022}.}  
  	Due to the compositional nature, the application of the chain rule,   gives a handy way to update the parameters when utilising gradient descent by what is known as the \emph{back-propagation} algorithm~\cite{werbos1975beyond}. In the following, we will denote the parameters of an MLP by $\Theta$, and the space to which it belongs to with $\mathcal{W}\ni\Theta$, without explicitly talking about the sequential structure of the functional mappings. The optimisation algorithm finds a point $\Theta_0\in\mathcal{W}$ which best approximates a target function by reducing a metric between the approximated function $\tilde{y}(\mathbf{x})$ and $y$ for all $(\mathbf{x},y)$ in the training dataset.

 \subsection{Universal approximation theorems: a bird's eye view} 
The range of functions that neural networks can approximate are encapsulated within the so-called universal approximation theorems (UATs)(see~\cite{HORNIK1989359,Cybenko1989,LESHNO1993861,VOIGTLAENDER202333} for instance). These theorems investigate the class of functions that MLPs with a certain activation can approximate with finite but arbitrary number of hidden nodes and/or layers. Most of them prove that the set of functions represented by a certain class of neural networks are \emph{dense} in the set of bounded continuous functions in the $n$-dimensional hypercube. 

The notion of dense sets in a larger superset is a generalisation of the properties of the rational numbers $\mathbb{Q}$ and real numbers $\mathbb{R}$, with $\mathbb{Q}$ being dense in $\mathbb{R}$. This means that for any $r\in\mathbb{R}$, we can find $q\in\mathbb{Q}$ such that $|r-q|<\epsilon$ for any arbitrarily small $\epsilon>0$. To generalise this notion to more intricate scenarios, one defines a metric, which measures a ``\emph{distance}" between two elements of the set. A dense set $\mathcal{D}$ in a superset $\mathcal{S}$, is one where for all elements $t\in\mathcal{S}$, we can find an element $a\in\mathcal{D}$ which is arbitrarily close with respect to a metric $d(s,a)<\epsilon$.
For UATs, this is generally the metric induced by the supremum norm: $\sup_{\mathbf{x}\in D} |\hat{y}(\mathbf{x})|$, which is the least upper bound of the absolute value of the function $\hat{y}$ in the domain $D$. For a parametrised function $\tilde{y}(\mathbf{x})=f(\Theta,\mathbf{x})$ and a target function $\hat{y}$, we have 
\begin{equation} 
	\label{eq:sup_met} 
	d(\tilde{y},\hat{y})=\sup_{\mathbf{x}\in D}|\tilde{y}(\mathbf{x})-\hat{y}(\mathbf{x})|\quad.
	\end{equation}  Loosely speaking, it measures the largest difference between the two functions in the domain $D$. Therefore, $\epsilon$ quantifies the highest absolute difference between the approximated function $\tilde{y}$ and the target function $\hat{y}$. 

The superset of interest for universal approximation is $C(I_n)$, the set of bounded continuous functions\footnote{Although, there are larger function spaces like the space of Lebesgue integrable functions $L_1(I_n)$ and possibility of UATs in them, we will restrict ourselves to $C(I_n)$ as finding effective training strategies for discontinuous functions are extremely difficult for the large dimensionality of modern neural networks.} from the closed unit hypercube $I_n=[0,1]^n$ to $\mathbb{R}$. 
One of the first universal approximation theorem~\cite{Cybenko1989}, proves that the set of functions represented by an MLP of one hidden layer and arbitrary but finite number of nodes with sigmoid activation is dense in $C(I_n)$ with respect to $d(\tilde{y},\hat{y})$.  This means that for any given target function $\hat{y}\in C(I_n)$,  there exists a  function $\tilde{y}(\mathbf{x})=f(\Theta_0,\mathbf{x})$ parametrised by an MLP of some finite number of nodes in the hidden layer and a point $\Theta_0$ in the parameter space which is an arbitrarily close approximation of $\hat{y}$. As an existence theorem, it does not talk about how to find the particular point $\Theta_0$, nor does it restrict it to any subset of $\mathbb{R}^N$, $N$ being the dimension of $\mathcal{W}$ which is the number of tunable parameters in the MLP. This dimension is controlled by the number of nodes in the hidden layer and the theorems do not provide a particular strategy to find $N$ for a target level of accuracy $\epsilon$. However, given the generality of the  function space which the theorem covers, the formal notion of existence of an arbitrarily close approximation to any bounded continuous function predicated the modern interest and revival of Artificial Neural Networks.

Practically, an MLP with a single layer is not a very efficient function approximator and falls within the broader class of shallow machine learning algorithms~\cite{scaling_in_book}.  This is because for highly fluctuating functions, one may need an exponentially large number of nodes compared to deeper architectures.   It was known intuitively that depth increases the effectiveness of  function approximation~\cite{scaling_in_book} which follows from close analogy to the computational complexity for width vs depth~\cite{10.5555/27031,10.1007/3-540-62034-6_33}  in circuits for implementing boolean operations, as well as the known ability of biological brains to learn simpler concepts first and more abstract concepts later in life. Deeper networks are able to approximate topologically more complex functions~\cite{6697897,pmlr-v49-eldan16} much more efficiently than shallow networks.\footnote{In ref.~\cite{6697897}, it is found that the sum of Betti numbers (a measure of topological complexity) of the subset of the domain of the function with positive outputs for a single output network scales at most as a polynomial in the number of nodes for several activation functions for a shallow network while it can scale exponentially for deeper networks.} Intuitively, this can be understood~\cite{JMLR:v21:20-345} from the ability of the component functions in the overall compositional chain to induce topologically discontinuous changes in each mapping.   

As the target function's closed form expression is not known, the approximation can only be as good as the quality and the amount available data. In HEP where synthetic data can be simulated based on well-understood theoretical formulations, the amount and quality of data  is seldom a problem. Yet the generality of the functions that dense architectures can approximate results in their handicap. The efficient optimisation via gradient descent algorithms only finds the best point $\Theta_0$ accessible from the initial point $\Theta_i$. However, domain knowledge from particle physics, motivate additional restrictions on the functions beyond continuity. Such restrictions can be built into the architecture which biases the functions to (strongly) always follow a particular property, or to (weakly) partially have some desired properties. We present a qualitative discussion of why such restrictions are needed in the next subsection.

\subsection{Restrictions beyond continuity: no such thing as a free lunch}
As we have seen hinted in the previous description, the existence of arbitrarily close approximation via a dense neural network to any given continuous function does not necessarily translate to finding the said approximation. Due to their non-constructive nature, one has to often revert to heuristics and brute force searches in the form of hyperparameter\footnote{The different choices given a base architecture like the activation function, optimisation method, learning rate, the number of hidden layers, the number of nodes in these layers etc., are known as hyperparameters.} optimisation coupled with some form of gradient descent optimisation to find a workable architecture and a particular set of parameters in the weight space. Although this is manageable for moderately large input dimensionality, it becomes intractable very fast with increasing number of dimensions. This is true even for deep fully-connected networks with more than two hidden layers. 
To counter this issue, additional informative priors  are generally employed when one goes to very high dimensional input spaces which restricts the function beyond continuity. In this subsection, we discuss the motivation and the need to assume these priors.

It is often enticing to think that given the universal nature of dense neural networks, an equally universal algorithm could exists that would guarantee an efficient optimisation for all distributions. However, this is known to be untrue~\cite{Wolpert1992OnTC,6795940,SCHAFFER1994259,585893,Wolpert2021}, where it was found that without any assumption about the underlying distribution, there is no such thing as an algorithm which performs the best in comparison to any other algorithm over the full range of possible functions. A particularly relevant exposition for the choice of learning algorithms and the inherent balance in its performance in all possible learning tasks can be found in ref.~\cite{SCHAFFER1994259},\footnote{Also, see ref.~\cite{scaling_in_book} for a discussion on the quest for artificial intelligence and its interplay with priors and these \emph{no-free-lunch} theorems.} where a conservation law for off-training generalisation error was proved over the full range of target functions for a fixed input distribution and, finite and fixed training size. \emph{Simply put, an algorithm performing well for a set of target functions will lose it's performance and be worse than a random guess in a different set of target functions so that the overall sum balances over the full space of target functions.} Therefore, if one knows some properties of the target function beforehand, it is possible (or mandatory perhaps) to devise algorithms or models which favour that particular form. This motivates building restrictions in the  possible functions that a non-parametric model (neural networks for our case) can approximate.

The restriction of the possible functions beyond continuity can be done primarily in three ways: the data representation, the architecture of the network, and the loss function including the choice of regulariser and the training procedure. The first two are related although for a given data representation, there can be competing choices of architectures based on the complexity. The last strategy can be regarded as independent of the first two while depending more on the nature of finding the optimal weights for a given data representation and architecture like convex vs non-convex optimisation, or saddle point finding vs extrema finding of the loss function. Strictly speaking, \emph{depth} is also one such prior which forces the learning of hierarchical abstractions of the input data relevant to the particular function that needs to be learnt at the output layer. However, with increasing input dimensionality, this alone is not enough to guarantee efficient search of the target function and one has to include domain-specific constraints into the architecture. For instance, the structure of convolutional neural networks with local connectivity, parameter sharing etc., is motivated from the fact that images tend to have local features which are relevant for (say) classification and they can be present anywhere in the image.

When the architecture is structured according to domain-specific priors, there is usually an additional fully-connected module which takes the features and process them for the final goal of the training. Therefore, the actual function that is approximated (by the downstream network) may not follow the implemented priors. The initial domain-specific modules are colloquially known as feature extractors, meaning, they replace the feature-engineering stage which builds domain-specific features normally utilised as inputs to shallow machine learning methods. Although deep-learning is a subset of machine learning algorithms, their distinction from shallow methods resides in their ability to find better functions than those relying on traditional hand-engineered features.  There is a long history of the usage of shallow machine learning methods~\cite{DENBY1988429,PhysRevLett.65.1321} in collider experiments and phenomenology (see ref.~\cite{Radovic2018} for a review). These methods utilise physics-motivated variable definitions constructed by processing the low-level high-dimensional data that one records at the detectors.  
Although, dense feed-forward neural networks with more than one-hidden layer, i.e., with no additional domain-specific designs, qualify as deep-learning models, the introduction of domain-specific design is what sets apart the modern algorithms which do not need extensively processed data. They can utilise the very high-dimensional low-level data and mostly outperform (dependent on the suitability of the priors) shallow machine learning methods.

For LHC phenomenology, all three methods of restricting the target function are utilised and they have been generally successful. However, due to the inherent nature of scientific research where favourable outcomes get published, there is a selection bias which can sometimes lead one to the false belief of out-of-the-box deep learning algorithms always performing better than shallow methods acting on physics-intuitive observables devised by experts. For instance, see refs.~\cite{Moore:2018lsr,Faucett:2020vbu} for a more pragmatic standpoint on Convolutional Neural Networks' comparison to substructure variables. Nevertheless, the wealth of prior physics knowledge also stipulates one to utilise it as much as possible while not being too restrictive to cause a performance degradation. Although the performance is only estimated \textit{a posteriori}, it has been found that physics-inspired networks generally help better the performance, interpretability, and training convergence. We will discuss these priors and their relation to the point cloud representation in the next section.

\section{Broad priors at the energy frontier}
\label{sec:hep_priors} 
The nature of collisions at high energies is described via the Standard Model (SM) of particle physics-- a Quantum Field Theory of the fundamental particles based on the gauge group $SU(3)_C\otimes SU(2)_L \otimes U(1)_Y$. Consequently, investigating particle collisions at the LHC has an inextricable connection to the underlying formalism, especially for phenomenological searches of new physics and particles. The nature of high-energy collisions result in the suitability of the point cloud representation to analyse events at the LHC. However, the restriction of the possible set of functions that neural networks can approximate based on priors is not restricted to physics, and most architectures in modern deep-learning usage inherently assume some form of symmetries or biases which help in the effective approximation for the particular domain. The underlying theme that unites the different deep learning models is the assumption of some form of geometry of the input data. With deep connections to functional analysis, differential geometry, and invariant theory, the study of the interplay between geometric priors within the architecture design, which restricts the function beyond continuity, resulting in efficient learning for various domains, is collectively referred to as geometric deep learning~\cite{7974879,bronstein2021geometric}. Consequently, other than building physics-inspired biases into the architecture, existing architectures with their own domain-specific priors have proved to be advantageous for the physics programme at the LHC.  In this section, we will discuss domain-specific priors resulting in the suitability of using the point cloud representation, as well as the compatibility of existing priors in the context of LHC phenomenology.  
\subsection{Physical Symmetries} 
\label{sec:symm} 
One of the most important priors that result from physics knowledge is the existence of symmetries. When training a neural network for a particular purpose, without any additional restriction, the network does not know of the already known symmetries like Lorentz invariance. These symmetries can be built into architecture using group equivariant neural networks, where the hidden representations follow the group multiplication property and the maps between them commute with the group actions on the underlying spaces. 

Given a group $\mathcal{G}$ and a set $\mathbf{X}$, the left-action of $\mathcal{G}$ on $\mathbf{X}$ is a map $A:\mathcal{G} \times \mathbf{X} \to \mathbf{X}$ with the following properties: 
\begin{enumerate}
	\item  $A(e,\mathbf{x})=\mathbf{x}$ for the identity element $e\in\mathcal{G}$ and any $\mathbf{x}\in\mathbf{X}$, and   
	\item  $A(g_1,A(g_2,\mathbf{x}))=A(g_1g_2,\mathbf{x})$  for all $g_1,g_2\in\mathcal{G}$. 
\end{enumerate}

 A map $f:\mathbf{X}\to\mathbf{Y}$ from sets $\mathbf{X}$ to $\mathbf{Y}$ which both admit group actions, say $\mathbf{A}_\mathbf{X}$ and $\mathbf{A}_\mathbf{Y}$ respectively, for the group $\mathcal{G}$ is said to be $\mathcal{G}$-equivariant if $f$ commutes with the group actions, i.e., 
 \begin{equation}
 	f(\mathbf{A}_\mathbf{X}(g,\mathbf{x}))=\mathbf{A}_\mathbf{Y}(g,f(\mathbf{x})) \quad,
 \end{equation}for every $g\in\mathcal{G}$ and $\mathbf{x}\in\mathbf{X}$.  The map $f$ is $\mathcal{G}$-invariant if the action $\mathbf{A}_\mathbf{Y}$ is trivial, i.e., $\mathbf{A}(g,\mathbf{y})=\mathbf{y}$ for all $g\in\mathcal{G}$ and any $\mathbf{y}\in \mathbf{Y}$.

Looking at the form of the group action, it is evident that it has a very close connection to  group representations. Given a representation of the group $\mathcal{G}$ in terms of invertible matrices of dimension $n\times n$,  $\rho: \mathcal{G}\to GL(n,\mathbb{R})$ which preserves the group multiplication in terms of the matrix multiplication (a group homomorphism), we can define an action on $\mathbb{R}^n$ as  $$A(g,\mathbf{x})=\rho(g)\,\mathbf{x}\quad,$$ where $\rho(g)$ are $n\times n$ invertible matrices. Consequently, the Lorentz transformation 
\begin{equation} 
	\label{eq:lorentz_action} 
	p'_\mu= \Lambda^\nu_\mu\,p_\mu\quad,
\end{equation}  is a linear action of the Lorentz group on the set of physical four-vectors $p_\mu$. Similar considerations hold for the multitude of groups that one encounters in high-energy physics, although the underlying field may at times be complex.   

Evidently the use of function approximators which already take into account the group structure reduces the search space of possible target functions immensely. This may\footnote{As the dimensionality of the input space is still very large, it is hard to ascertain the improvements \textit{a priori}.} result in an increase in sample efficiency (the number of training samples required to reach a particular level of performance), convergence (the number of epochs required to reach that particular accuracy), reduced number of trainable parameters, or increase in overall performance with respect to non-equivariant architectures. This gain is dependent on the suitability of the target function and the particular group equivariance that is implemented into the architecture that learns the target function.  As a simplified example, take the binary classification problem of two resonant particles with different (but with some overlap due to their width) masses  decaying into two particles. The mass of the resonance is obviously a very important quantity in the discrimination and a Lorentz invariant feature extraction from the sum of the decay products' four vectors would expose this feature directly, while, say Euclidean group invariance in eight dimensions would struggle to find the distinguishing feature. From a purely data driven perspective,  the four vector of the two particles are just some feature vector living in $\mathbb{R}^8$, and it is our physics intuition which drives the additional algebraic structure of the Lorentz group. 

Equivariant neural networks have been instrumental in several domains ranging from image processing~\cite{pmlr-v48-cohenc16,pmlr-v97-cohen19d}, to the prediction of protein structures~\cite{Jumper2021}. With the rich structure of symmetries in particle physics, it is no wonder that the community is exploring novel ways to utilise equivariant architectures~\cite{Kanwar:2020xzo,Favoni:2020reg,Bogatskiy:2020tje,Buhmann:2023pmh} for various applications in theory and phenomenology. 

\subsection{Permutation invariance and variable cardinality} 

Another important aspect of studying collisions at the LHC, is the inherent absence of order in the data. In event analysis, the reconstructed objects like jets, leptons, photons etc., have four vectors and additional information like flavour and charge which form an unordered collection. Similarly in jet substructure analysis, the ordering of the constituents are irrelevant while the discriminating information is found in the inter particle correlations.  Due to the non-conservation of particle multiplicity in relativistic quantum field theory where highly energetic particles inevitably produce additional particles which share the momentum, the number of elements in the set for either case is not constant. Therefore, one needs to consider an input representation which acts on unordered sets of variable cardinality, i.e., to preserve the property of sets, a function needs to be permutation invariant with respect to the order of the inputs, and be well-defined for varying number of inputs.  

Restricting events to an ordered and fixed sized representation may result in truncation and combinatorial ambiguities~\cite{Onyisi:2022hdh}.  While we may be interested in a specific number of coloured objects at the parton level, due to their highly radiative nature, it is often advantageous to not have a hard restriction on the number of reconstructed jets with additional jets arising from additional radiations. In other words, truncating an event signature to a fixed sized representation introduces ambiguities in how one should select these fixed number of objects with the kinematics inextricably linked to those objects which are not utilised in the representation.  Combinatorial ambiguities arise when a signature has more than one way of associating the reconstructed objects to an earlier parton. For instance, in the fully leptonic decay of a pair of top quarks, other than the inability to reconstruct the momentum of the two neutrinos, we also have to decide the allocation of two bottom jets (as we cannot determine the charge) to the two reconstructed lepton. The situation becomes more complicated if we have more than two identified bottom jets in a candidate event. 

Truncation ambiguities are much more pronounced in jet-substructure studies, where analogues of multiplicity are well-defined with more stringent theoretical considerations~\cite{Frye:2017yrw,Medves:2022ccw,Medves:2022uii}. This difficulty is due to the resolution of QCD radiation patterns at much smaller angular scales within the jet in the regimes with considerable collinear enhancement. Combinatorial ambiguities can also arise in signal decays into more than two particles, like the top quark, when one is interested in assigning a parton-level flavour to some of the constituent subjets.   
\subsection{Infrared and collinear safety: resilience to deformations} 
The presence of massless particles, demand an additional consideration for perturbative calculations in QFT, namely, infrared and collinear (IRC) safety. This is one of the cornerstones for fixed order calculability of any observable for theories containing massless gauge bosons, where there is an intricate cancellation of infrared divergences arising from virtual loops with the interference contribution of soft and collinear emissions. This cancellation is guaranteed to happen for IRC safe observables, i.e., when an observable defined for $n+1$ particles is equal (or as a limit) to the same for $n$ particles when any of the $n+1$ particles become soft or any two of them become collinear. Therefore, IRC safety puts additional relations between the functions defined on sets with cardinality difference of one.   

In domains like image processing or 3D point cloud processing, neural networks have some form of pooling or readout operation in a local region which propagates the summary statistic of the region forward. This results in the propagated information being approximately resilient to \emph{local deformations} in the data, which mainly arise from noise. Due to the enhancement of radiation in the almost collinear or soft regimes, there is an inherent deformation of the data living in the rapidity-azimuth plane for the point cloud or the image representation.  Near collinear emissions result in local distortions in a \emph{jet-image}, while soft emissions are inherently less prominent since the harder hits will dominate the sum in the convolution operation. Therefore, Convolutional Neural Networks already take into account resilience to such soft and collinear emissions~\cite{Choi:2018dag}. On the other hand, point cloud architectures which take the three dimensional features $(p_T,\Delta \eta,\Delta \phi)$ of each particle as inputs to an MLP, do not distinguish between the transverse momentum and the angular variables and as such, the extracted features are IRC unsafe. Nevertheless, the very general nature of point cloud representation allows defining IRC safe feature extractors acting on sets of particles as epitomised by Energy Flow Networks~\cite{Komiske:2018cqr,Dolan:2020qkr,Shen:2023ofd,Bright-Thonney:2023gdl,Gambhir:2024dtf} along with its graph and hypergraph generalisations~\cite{Konar:2021zdg,Atkinson:2022uzb,Konar:2023ptv,Chatterjee:2024pbp,Bhardwaj:2024djv}. 

An important aspect of the local deformations induced by QCD radiation is in its inevitability even in the assumption of perfect measurements, since  it results from the underlying physical process rather than being induced by measurement errors. Therefore, IRC safety is a crucial element of any phenomenological analysis at hadron colliders. While the IRC safe anti-$k_t$~\cite{Cacciari:2008gp} jet definition has been the main workhorse of physics analyses at LHC, jet definitions utilised at Tevatron were still IRC unsafe~\cite{Andrieu:2005re,Salam:2010nqg}. One should note that theoretical predictions even between IRC safe jet definitions can result in systematic differences of 10\% or larger as shown in ref.~\cite{Ellis:2009zyy}, and without IRC safety, we are at the mercy of poorly understood non-perturbative effects. As deep learning approaches to event analysis will presumably expand in the future, it is important to consider their theoretical aspects within perturbative QCD. In this regard, the algorithm's IRC safety is an important requirement for its understanding, especially since IRC-safe deep learning algorithms do not restrict the expressivity to a very high degree for jet-tagging.\footnote{If the information of interest for a particular training objective is in the high-energy patterns in an event, one would intuitively expect an IRC-safe algorithm to be relatively comparable to unsafe ones, as IRC-safe observables concentrate on the hard energy flow of the event.} However, IRC safety alone does not equate to the perturbative physics being under control, or the absence of non-trivial non-perturbative effects, and additional considerations like the algorithm's all-order effects~\cite{Bright-Thonney:2023gdl, Banfi:2004yd,Dasgupta:2013ihk,Kasieczka:2020nyd} need to be properly addressed, with little loss (preferably without losing out) on their generalisation capabilities. On the other hand, there are also classes of observables~\cite{Chang:2013rca, Larkoski:2015lea} that do not require fixed-order IRC safety but can be studied with more elaborate theoretical underpinnings. Notably, additional sources of deformation also arise from experimental measurements, as well as the presence of pileup interactions.      

\subsection{Priors from image processing} 
Convolutional Neural Networks (CNNs)~\cite{FUKUSHIMA1982455,6795724} remain the quintessential example of deep learning algorithms with very good generalisation abilities after curtailing the expressivity based on domain knowledge. On top of being approximately resilient to local deformations as explained in the previous subsection, they also assume local connectivity and parameter sharing, which, when combined with successive composition of layers result in scale separation. These biases are carried over in some sense into graph-based architectures depending on specific details of the architecture which we will touch upon in the next section. Here, we discuss the natural ways in which these three assumptions already account for the structure of QCD radiation patterns arising at the LHC. 

 \begin{figure}[t]
	\centering 
	\includegraphics[scale=.3]{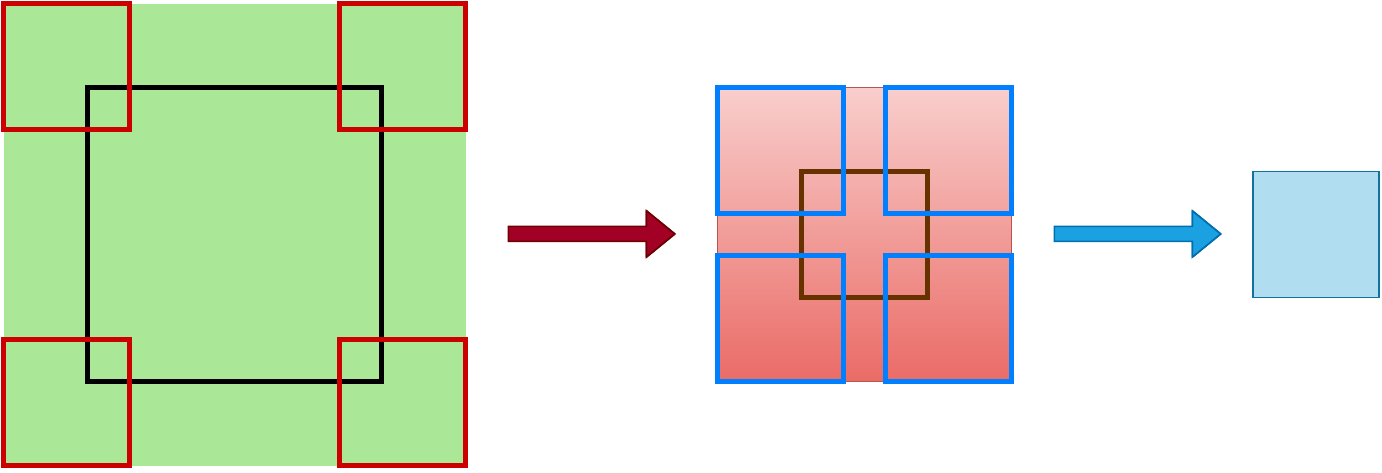}
	\caption{The figure shows a diagrammatic representation of the effective region (shown as a solid greed box) in the input image for a second convolution on the first feature map. The result of the convolution on the green region of the image with the filters (shown as hollow red squares) produces a part of the feature map shown as a solid red box. 
		Similarly, a second convolution on the region of the feature map with the different filters of the same size produces a region of a new feature map. Therefore, while the first filters look into the local region corresponding to their actual size, the second filters learn the features in the image corresponding to a much larger area (determined by the relative size of the filters) in the original image.} 
	\label{fig:scale_sep} 
\end{figure}

\begin{figure}[t]
	\centering 
	\includegraphics[scale=0.5]{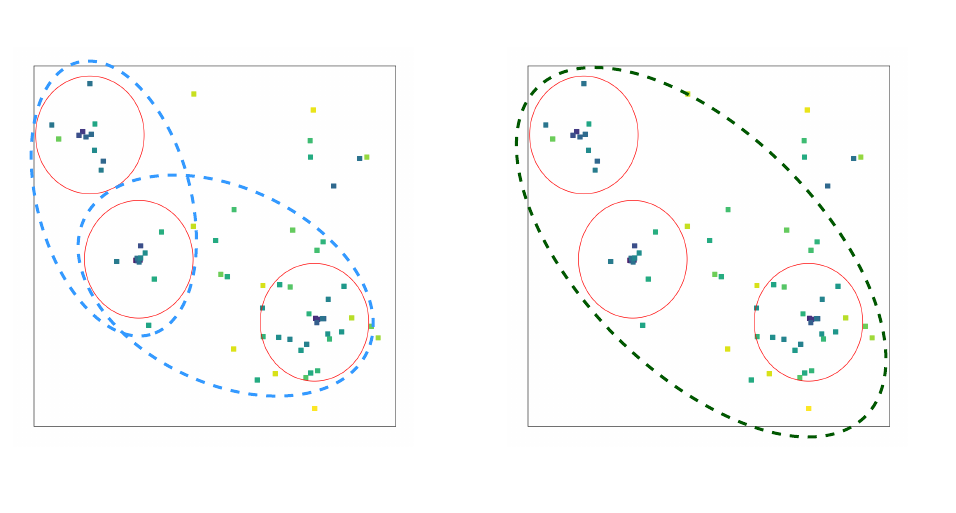}
	\caption{The hierarchical separation of scales encountered in a top jet image with the little boxes denoting calorimeter hits. The red circles denote three subjets, which represent the first local features in the jet. On the left, the blue ovals denote the scale at which the two prong structure becomes visible, while on the right the green dotted oval is the scale at which all three subjets become relevant.} 
	\label{fig:top_eg} 
\end{figure}

Local connectivity means that for a given input pixel or a node, the information processed by the network concentrates in its local neighbourhood. Compared to the image size, the smaller  filters in a CNN act on local regions determined by their relative sizes. The collinear splitting structure in parton showers which result in the formation of jets is an inherently local phenomenon in the rapidity-azimuth plane. Therefore, the local nature of feature extraction looks into the local substructure in jets or jets in events. Parameter sharing refers to reusing the same set of weights for different local regions in the input. In QCD, the universality of parton showers results in similar patterns of radiation over the full detector\footnote{Even though pile-up break this generality, this remains true after pileup removal after which phenomenological analyses are conducted on events following such patterns.} in different positions which motivates reusing the same set of weights to pick up such features. Separation of scales occurs via successive composition of functions acting on local regions in the input representation. Being a more refined version of depth to learn hierarchical abstractions of the data, the locally acting functions create a separation of scales, where the filters in initial layers look at the immediate locality and deeper ones sequentially capture larger scales. This is diagrammatically explained in figure~\ref{fig:scale_sep}. Consequently, hard prongs within jets or jets within events are present at different scales in the rapidity-azimuth plane, and a separation of scales help look into these structures in a hierarchical manner. An example for a three prong top jet is shown in figure~\ref{fig:top_eg}.

\section{Graph based architectures} 
\label{sec:gnn} 
In the previous section, we discussed some pertinent characteristics of high-energy collisions at LHC that motivate a set-based representation of events. In this section, we elaborate on the set-based representation and architectures that preserve the set property, concentrating on the graph representation that endows relational structures between pairs of constituents of the set.

\subsection{Functions on sets} 
The set-based representation where the elements of the set represent some feature of objects in any underlying metric space is known as point clouds. It originates from the set-based representation of discrete points in 3D coordinate space for use in various computer graphics and imaging applications while being general enough to allow the elements to reside in more abstract spaces like those found in high-energy physics.

As we are interested in sets of finite but arbitrary size consisting of features living in an abstract space, say $\mathcal{X}$, we are generally interested in learning functions from the power set\footnote{If the cardinality of $\mathcal{X}$ is infinity, we are interested in elements of the power set which has bounded cardinality.} of $\mathcal{X}$ to the output domain $\mathcal{Y}$. Such functions are required to be permutation invariant with respect to the input ordering to preserve the property of sets as unordered collection of objects. The universality of such functions was studied in the Deep Sets \cite{NIPS2017_f22e4747} framework, which defined set based feature extractors that posses the property of permutation invariance or equivariance. For the invariant case, a Deep Sets model $f$ is of the form 
\begin{equation}
	f(\{\mathbf{h}_1,\mathbf{h}_2,.....,\mathbf{h}_M\})=\rho\left(\sum_{i=1}^{M}\,\phi(\mathbf{h}_i)\right)\quad,
\end{equation}
where $\mathbf{h}_i$ are the set elements, $M$ the cardinality of the set, and $\rho$ and $\phi$ are MLPs.  We can see that the function $\phi$ learns a per-object map to a latent space, which undergoes a permutation-invariant summation over the set constituents. Going into the two-part segregation of deep learning algorithms, the MLP $\phi$ acts as a feature extractor while $\rho$ gives the downstream output. The extracted features are combined in a permutation invariant manner under the summation, which reflects the prior of set-based functions. Note that the summation can be replaced by any permutation invariant operation like taking the component-wise maximum or minimum over the latent representation.

Although the Deep Sets algorithm is a powerful feature extractor, the feature extraction stage looks only into the individual elements of the set. The downstream MLP, on the other hand, has access to the aggregated global information of the entire set. Such an architecture does not exploit the relational information between the set's constituents. To circumvent this issue, one endows additional structures on the set, like graphs, which expose pairwise relationships, or hypergraphs, which generalise graphs to account for any $n$-element relationships within the set.

\subsection{Describing data with graphs} 
As we have describe above, graphs equip the input set with relational information between two elements. These elements in graph terminology are called nodes or vertices. The edge set $\mathcal{E}$, is a subset of the Cartesian product $\mathcal{S}\times\mathcal{S}$. The presence or absence of an edge, say $(i,j)$ in $\mathcal{E}$ is determined by usefulness of the relation between the two elements $i$ and $j$. Consequently, the way a graph is constructed determines the information highlighted amongst the nodes. In LHC physics, these may include forming edges between geometrically close components in a detector module, a Feynman diagram based construction, or closeness in the rapidity-azimuth plane. 
If $\mathcal{E}=\mathcal{S}\times\mathcal{S}$, the corresponding graph is called a complete graph, the relation between every pair of elements is exposed in the graph. This is generally avoided as it leads to a quadratic scaling with the input set cardinality.

A graph can be represented as $\mathcal{G}=(\mathcal{S},\mathcal{E})$, where $\mathcal{S}$ represents the set of nodes and $\mathcal{E}$ represents the set of edges. Each element in the set is some feature vector $\mathbf{h}_i\in\mathbb{R}^d$. However, the set $\mathcal{S}$ may at times refer to the index set consisting of $i\in\{1,2,....,|\mathcal{S}|\}$ in the following discussions which will be clear from context. The edge set $\mathcal{E}$ for computational purposes, can be suitably represented as an $|\mathcal{S}|\times|\mathcal{S}|$  matrix $\mathbf{A}$, with components
\begin{equation*}
	\label{eq:adj_mat}
	A_{ij}= \left\{\begin{array}{cc}
		1\quad & \text{if } (i,j)\in\mathcal{E}\\
		0\quad & \text{otherwise} 
	\end{array}\right.\quad,
\end{equation*}  which is called the adjacency matrix of the graph. Note that $\mathbf{A}$ assumes a particular ordering of the nodes. This is the same when we input the graph quantities to a computer in the form of an array. Although there can be complexity studies of graphs where the nodes are considered indistinguishable, the nodes for our case are distinguished by their feature vectors $\mathbf{h}_i$. Moreover, the edges for our case will be directed as the message functions (to be explained later) will generally be asymmetric to interchange of the two constituent nodes of an edge.  

The addition of edges as relational pairs require the conservation of the overall structure of the edge set for any permutation of the nodes. This is a stricter requirement than permutation invariance of the nodes, and formally covered within the notion of \textit{graph isomorphisms}. Any two graphs $\mathcal{G}=(\mathcal{S},\mathcal{E})$ and  $\mathcal{G}'=(\mathcal{S}',\mathcal{E}')$ are isomorphic to each other if we can find an invertible function $N:\mathcal{S}\to\mathcal{S}'$ between the node sets which induces an invertible map $E:\mathcal{E}\to\mathcal{E}'$ between the edge sets. As permutation of the elements of $\mathcal{S}$ can be viewed as a bijection of the set to itself, graph  isomorphisms additionally require the conservation of the edge structure for any permutation of the nodes. Although the notion of graph isomorphisms is often equated to being oblivious to relabelling of the nodes, our case is not as straightforward as assuming such a simple relabelling as the node features are immutable intrinsic labels of the nodes. On top of these node features, we assign an index to the the array containing each node features and construct the edges as a two-dimensional arrays of size $2\times|\mathcal{E}|$ in the sparse representation, or the adjacency matrix of size $|\mathcal{S}|\times|\mathcal{S}|$. Graph isomorphisms then practically relate to how the node features are arranged in this representation with the graph property intact as long as we work with quantities which are invariant for different ordering of the node features.

\begin{figure}
	\centering
	\includegraphics[scale=0.23]{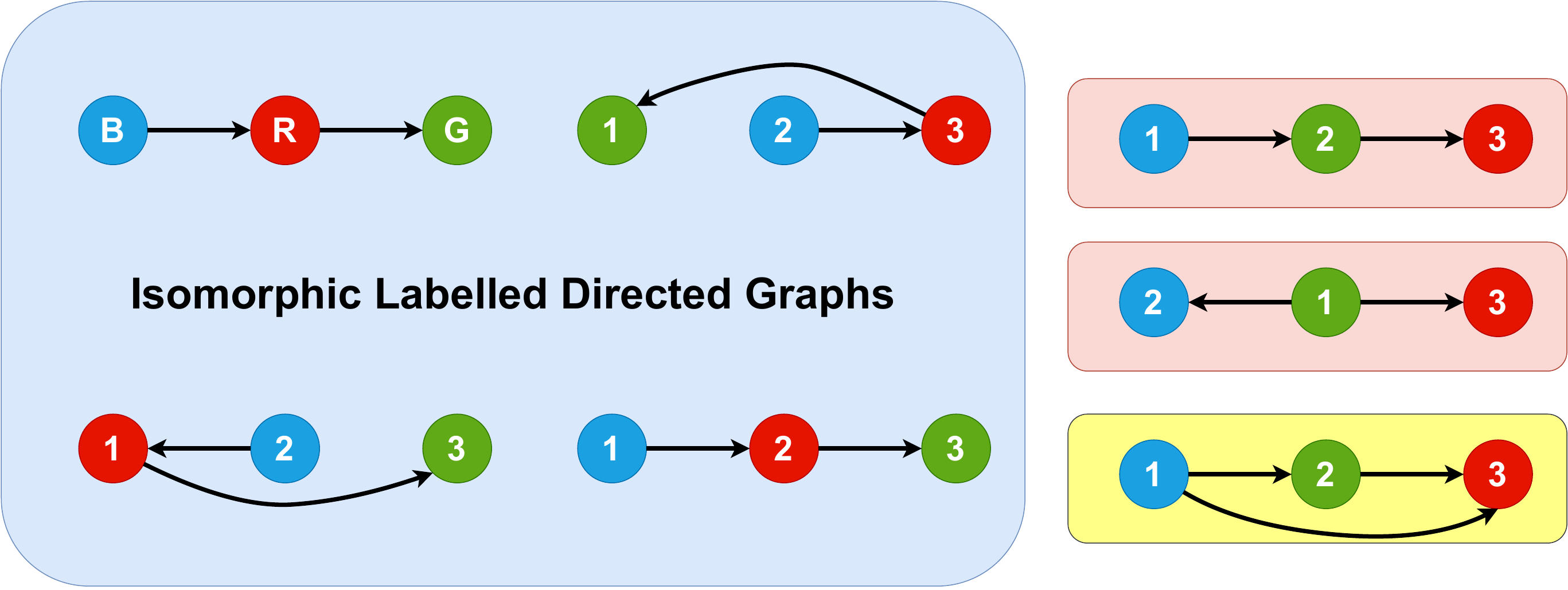}
	\caption{Examples of isomorphic and non-isomorphic graphs between directed graphs with nodes distinguished by their colours. The graphs in the bounded blue box are isomorphic to one another while the three on the right are not isomorphic to any other graph. The two graphs within the red boxes become isomorphic to the blue ones when one forgoes the difference in the colours and the direction of the edges. }
	\label{fig:graph_iso} 
\end{figure}

A representation of different isomorphic and non-isomorphic graphs with three coloured nodes is shown in figure~\ref{fig:graph_iso}. The colour in our case represents the node features which are intrinsic to that particular node. Therefore, one has to take into account this intrinsic character when defining graph isomorphisms. All graphs within the blue bounded box are isomorphic to each other as one can easily define invertible maps between the node labels which preserved the underlying directed connections between the coloured nodes. We have intentionally ordered the nodes in a straight line and numbered them in increasing natural numbers (except for the one on the top-left) to make it evident the analogy to a matrix based representation where the numbers denote its index. On the right all three graphs are not isomorphic to each other or to any of the ones in the blue box when considering the colours and edge directions. As there can be no bijections between node sets or edge sets of different sizes, the graph on the yellow box at the bottom can never be isomorphic to any other graph in the figure when relaxing the additional constraints of colourings and directions. However, the graph on top right is isomorphic to the ones in the blue box if one forgoes the colourings but keeps the edge directions. This is not the case for the one in the middle since the green node has two outgoing edges while it is not the case in the other graphs. Finally, all graphs except for the one in the yellow box are isomorphic to each other if one forgoes both the colourings and the edge directions. 

We now explain some terminologies which will be used in the following discussion of graph neural networks. The neighbourhood of a node $i$, written as $\mathcal{N}(i)$ is the set of all connected nodes to $i$. For directed graphs, one usually regards either the incoming or outgoing edges (but not both) to define the neighbourhood.  For instance, in the isomorphic graphs within the blue box in figure~\ref{fig:graph_iso}, the neighbourhood of the red node if one considers incoming edges is made up of the blue node, while for outgoing connections it is the green node alone. The $l$-hop neighbourhood of a node $i$ consists of all nodes which can be reached from the node $i$ by following at most $l$ edges. Like before, the direction should be consistently defined for directed graphs and $l=1$ results in the usual neighbourhood definition. In figure~\ref{fig:graph_iso}, considering outgoing edges, the blue node's  2-hop neighbourhood consist of the green node on top of the red one. For our discussion of GNNs, we will consider the neighbourhood $\mathcal{N}(i)$ to be defined as the set of nodes with incoming connections to $i$. Thereby, we will denote any edge attribute connecting two nodes of indices $i$ and $j$ with the subscript $i\leftarrow j$ to make the direction self-evident.


\subsection{Graph Neural Networks}
In order to extract features from a graph for graph-level purposes, we need an architecture which extracts features which are invariant with respect to graph isomorphisms\footnote{Technically, this means that the graph representation is defined for an equivalence class of isomorphic graphs rather than each individual graphs in the class.} where the extracted features are identical for any two isomorphic graphs. For node or edge level applications where one uses their features rather than the global graph representation, the output of the network should depend only on the intrinsic features and not on the index of the features, i.e., for any two isomorphic graphs we can obtain an invertible map between the node sets which preserve the updated features of the graph components.   

One of the first proposals of Graph Neural Networks~\cite{1555942,4700287} (GNNs) for any general graph compatible with back-propagation optimisation devised the concept of states of nodes and edges adapting their usage in recursive neural networks~\cite{572108,712151} which acts on directed acyclic\footnote{An acyclic graph is one where there are no paths with distinct nodes between any two nodes which form a closed loop. } graphs. The essential quality which conserved the graph properties  consisted of recursively updating the states in successive time steps with a shared parametric model like an MLP. More recent generalisations include the Message Passing Neural Network (MPNN)~\cite{pmlr-v70-gilmer17a} approach and the more general Graph Networks~\cite{Battaglia2018RelationalIB}. Our discussions will concentrate on the more popular MPNNs.  

An MPNN consists of several message-passing operations that takes in a graph $\mathcal{G}=(\mathcal{S},\mathcal{E})$ with node features $\mathbf{h}_i\in\mathcal{S}$, and optionally, edge features for all $(i,j)\in\mathcal{E}$ as $\mathbf{e}_{i\leftarrow j}$ and updates the node features via the following steps:

\begin{enumerate}
	\item 
\textbf{Message passing:} A learnable function (generally an MLP) $\phi_e$, called the message function takes the node features $\mathbf{h}_i$ and $\mathbf{h}_j$, for each edge $(i, j)\in\mathcal{E}$  and optionally the edge feature $\mathbf{e}_{i\leftarrow j}$ and evaluates the message as 
	\begin{equation}
	\label{eq:msg_pass}
	\mathbf{m}_{i\leftarrow j}=\phi_e(\mathbf{h}_i,\mathbf{h}_j,\mathbf{e}_{i\leftarrow j})\quad.
	\end{equation}
 \item \textbf{Message Aggregation:} For each node $i$, we aggregate the messages $\mathbf{m}_{i\leftarrow j}$ for all nodes $j$ in the neighbourhood $\mathcal{N}(i)$ with a permutation invariant function $\Box_l$  
	\begin{equation}
	\label{eq:node_readout}
	\mathbf{m}_i=\Box_l(\{\mathbf{m}_{i\leftarrow j}\,|\, j\in\mathcal{N}(i)\})\quad.
	\end{equation} 
\item \textbf{Node update:} The node features are updated to $\mathbf{h}'_i$ as a function $\phi_h$ (which also can be an MLP) of the aggregated message $\mathbf{m}_i$ and the input node feature $\mathbf{h}_i$ as 
\begin{equation}
	\mathbf{h}'_i=\phi_h(\mathbf{h}_i,\mathbf{m}_i)
\end{equation}
\end{enumerate}

The number of message passing operation is fixed for a particular model and one either uses the messages $\mathbf{m}_{i\leftarrow j}$ and the node feature $\mathbf{h}_i$ as inputs to an MLP for edge and node specific tasks, respectively. For graph level purposes, we constructs the graph representation $\mathbf{G}$, by a readout operation on the node features with an analogous global readout function $\Box_g$ as $\mathbf{G}=\Box_g(\{ \mathbf{h}_i | i\in \mathcal{S} \})$ .

It is now evident that an MPNN's local connectivity is determined by the graph's structure with the first message-passing operation looking at the immediate neighbourhood, and successive ones looking at larger and larger $l$-hop neighbourhoods. The parameter sharing (the message function or the node update MLP is the same for all edges and nodes), and the message- aggregation stage, brings forward the biases discussed for CNNs to GNNs. However, while the pooling operations downsample the image representation thereby reducing redundant information from going deeper into the architecture, the readout operation does not reduce the cardinality of the node set and there is a much faster increase of redundant information amongst the node representations as one goes deeper. This hugely restricts the practical viability of deeper MPNNs for small graphs as the node features become increasingly similar with more message-passing operations.

The GN formalism additionally consists of edge updates and global feature updates, however, most common GNNs utilised in LHC phenomenology can be explained as a specific form of the MPNN approach.
For instance, the message function in edge-convolutions~\cite{10.1145/3326362} is of the form  
\begin{equation*}
	\mathbf{m}_{i\leftarrow j}=\phi_e(\mathbf{h}_i\oplus \mathbf{h}_j-\mathbf{h}_i) 
	\end{equation*} with the node updated via $\mathbf{h}'_i=\mathbf{m}_i$ after a choice of permutation invariant aggregation. The operation $\oplus$ denotes concatenation of the two vectors. 

Another important operation is the attention mechanism which currently have state-of-the-art performance on the public top-tagging dataset.   
Although attention mechanisms originated~\cite{DBLP:journals/corr/BahdanauCB14} as a way to circumvent the loss of information in Recurrent Neural Networks for natural language processing (NLP), the Transformer model~\cite{NIPS2017_3f5ee243} which solved the bottleneck for efficient training of sequential language models by utilising attention alone while forgoing the recurrence, also found state-of-the-art performance for various NLP tasks and opened the proverbial ``Pandora's box" of Large Language Models~\cite{devlin-etal-2019-bert,NEURIPS2020_1457c0d6,NEURIPS2022_b1efde53,radford2018improving}. While these exciting developments are outside the scope of the current review, the attention mechanism being a set operation is applicable to point clouds~\cite{gat,9710703}. The message function for any general attention mechanism can be written as $$\mathbf{m}_{i\leftarrow j}=\mathbf{w}(\mathbf{h}_i,\mathbf{h}_j)\, \alpha(\mathbf{h}_i)  \quad,$$ for some learnable function $\alpha$ and attention mechanism $\mathbf{w}$ whose output dimension should either be a scalar or a vector with the same dimensions as that of $\alpha(\mathbf{h}_i$) for component-wise multiplication. The learnable function $\mathbf{w}(\mathbf{h}_i,\mathbf{h}_j)$ denote the attention of the node $j$ with respect to the source node $i$, and can be cast in different ways according the particular attention mechanism. For the interpretation of $\mathbf{w}(\mathbf{h}_i,\mathbf{h}_j)$ as weights in the summed aggregation $\mathbf{h}'_i=\sum_{j\in\mathcal{N}(i)} \mathbf{m}_{i\leftarrow j}$, the requirement $\sum_{j\in\mathcal{N}(i)} \mathbf{w}(\mathbf{h}_i,\mathbf{h}_j)=1$ is achieved by a softmax normalisation on the $j$ index. Due to this learnable importance of the different nodes, attention-based networks often employ a complete graph structure where the job is left to the attention mechanism to learn the respective inter-node importance.

\section{ Overview of applications in high energy physics} 
\label{sec:overview} 
Graph Neural Networks in their different guises have been explored in a varied range of applications for the physics program at the Large Hadron Collider. There can be several ways to categorise these applications based on the learning algorithm, the prediction task, and the physics application. The categories of learning methods broadly fall under unsupervised, semi-supervised and supervised methods, while the prediction task can be segregated into node, edge and graph level tasks. In this section, we give a brief overview of the applications of graph-based for some physics applications at the LHC.

\paragraph{Jet Classification:} The prototypical example of deep learning applications is the classification of large radius QCD jets~\cite{deOliveira:2015xxd} from boosted hadronic decays of heavy particles and various point-cloud architectures have been studied for jet classification. Carrying forward the point cloud representation, ParticleNet \cite{Qu:2019gqs}, which utilized dynamic graph convolutions, presented the first application and showed high efficiency in classifying  QCD/top and quark/gluon tagging. Interaction networks-based methods \cite{Moreno:2019bmu,Moreno:2019neq} were also explored for jet tagging for 2-prong and 3-prong boosted jets. Attention-based GNNs \cite{Mikuni:2020wpr} were explored for quark-gluon tagging applications. The point transformer~\cite{9710703} architecture has also been explored~\cite{He:2023cfc} for discriminating quark/gluon and top/qcd jets. Chebyshev graph convolutions~\cite{NEURIPS2022_2f9b3ee2} which enables the model to capture local dependencies intrinsic to jet formation,  have also been studied for jet-tagging~\cite{Semlani:2023kzf}. Haar pooling message passing networks~\cite{pmlr-v119-wang20m,Ma:2022bvt}, have also shown to improve tagging performance compared to the usual readout operations.  
Till date Particle Transformer~\cite{pmlr-v162-qu22b}, based on a modified multi-head attention mechanism~\cite{NIPS2017_3f5ee243,shleifer2022normformer}, shows state-of-the-art performance in the public top tagging dataset~\cite{Mikuni:2021pou}.

As already explained in Section~\ref{sec:symm}, physical symmetry has been used in graph-based architectures. Lorentz Group Network (LGN) \cite{Bogatskiy:2020tje} utilise the isomorphism of proper orthochronous Lorentz group SO$(1, 3)^+$ to the projective special complex linear group $\text{PSL}(2, \mathbb{C})$ to build tensorial representations decomposed into irreducible representations in terms of the Clebsch-Gordan coefficients. While LorentzNet \cite{Gong:2022lye,Li:2022xfc} implemented a simplified scalar and vector representation in the hidden representations to extract the features. Later, PELICAN \cite{Bogatskiy:2022czk, Bogatskiy:2023nnw} combined permutation and Lorentz equivariant aggregations. On the public top tagging data set, these architectures matched or outperformed ParticleNet with fewer trainable parameters. Meanwhile, LorentzNet showed a very high sample efficiency, reaching an AUC of 0.9839 with only $5\%$ of the training data. LorentzNet was augmented with capsule network~\cite{Sahu:2024sts}, showing improved performance for quark-gluon performance. The gain in performance comes because the feature extraction specifically utilises the physical symmetry of the dataset. 

When utilising the jet's constituents when extracting features with deep-learning models, it is not necessary that automatic feature extraction on a jet is IRC safe, and the IRC safety of these features depend not only on the input features but also on how they are structured. Constructing infrared and collinear-safe feature extractors on point clouds has also received considerable effort. 
 Energy Flow Networks~\cite{Komiske:2018cqr} (EFNs) adapted the Deep Sets\cite{NIPS2017_f22e4747} framework to account for IRC safety by utilising per-particle maps of the angular coordinates and performing a summed readout over the particles after linearly weighting with energy. Extensions of this approach include a permutation equivariant EFN~\cite{Dolan:2020qkr}, building hierarchical EFNs by utilising a basis of Legendre polynomials~\cite{Shen:2023ofd}, Lipschitz-EFNs~\cite{Bright-Thonney:2023gdl} where the extracted features are made to follow the Lipschitz continuity condition motivated from the geometry of particle collisions~\cite{Komiske:2020qhg}, and utilising higher moments~\cite{Gambhir:2024dtf}. The per-particle nature of EFN feature extraction has been generalised to graphs in the framework of Energy-weighted Message Passing Networks~\cite{Konar:2021zdg} for supervised classification. Further extensions include hypergraph based feature extraction~\cite{Konar:2023ptv}, and equivariance in the rapidity-azimuth plane~\cite{Chatterjee:2024pbp,Bhardwaj:2024djv}. An architecture based on dynamic edge convolutions was also studied for tagging jets originating from dark showers~\cite{Bernreuther:2020vhm}. 

Another theoretically motivated way to represent a jet as a graph is through the Lund tree de-clustering~\cite{Andersson1989,Dreyer:2018nbf}, which maps emissions according to the jet clustering history. LundNet~\cite{Dreyer:2020brq} employed Lund graphs as inputs for various binary jet classification tasks. Due to the theoretically transparent nature of the input construction, it allows more detailed semi-analytic studies~\cite{Dreyer:2021hhr} of its discriminative power than is possible with the other graph-based approaches.

 \paragraph{Discovering New Physics via Event Classification:} 
 
Graph Neural Networks (GNNs) can categorize particle collision events into different classes, distinguishing between signal and background events. Leveraging the inherent graph structure of particle interactions, GNNs are adept at capturing nuanced characteristics that conventional approaches may overlook, thereby improving the accuracy of event classification. GNNs are advantageous for the event classification task for LHC data as they overcome the shortcomings of CNNs, which are limited to the Euclidean domain. An event at a collider can be naturally represented as graphs, where the final state particle represents the node and their interactions are encoded by edges. 

\begin{figure}
	\centering
	\includegraphics[scale=0.5]{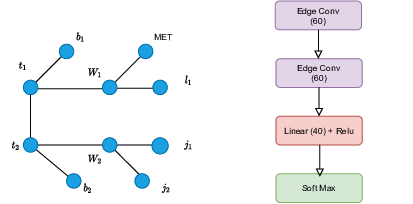}
	\caption{Example of a decay topology inspired graph (figure taken from ref.~\cite{Atkinson:2021jnj}). }
	\label{fig:dectop} 
	
\end{figure}

In phenomenological studies, fully connected graphs and physics-based topology-inspired graphs are explored for event classification. Physics-inspired topology-based graphs are an efficient approach to capturing complex data relationships while minimizing NN parameters. In a multiclass set-up for the semileptonic $t\bar{t}$ final states, GNN was used to discriminate the thirteen independent Wilson coefficients  \cite{Atkinson:2021jnj}. The decay-topology-inspired graph is shown in figure \ref{fig:dectop}. To construct the graph, final state reconstructed particles are first added as the nodes, and additional nodes are added to the topology if the parent of the decay product can be reconstructed.
On the other hand, fully connected graphs are also advantageous in case of unknown underlying physics topology or if fewer final state particles are involved in an event.

\begin{figure}
	\centering
	\includegraphics[scale=0.4]{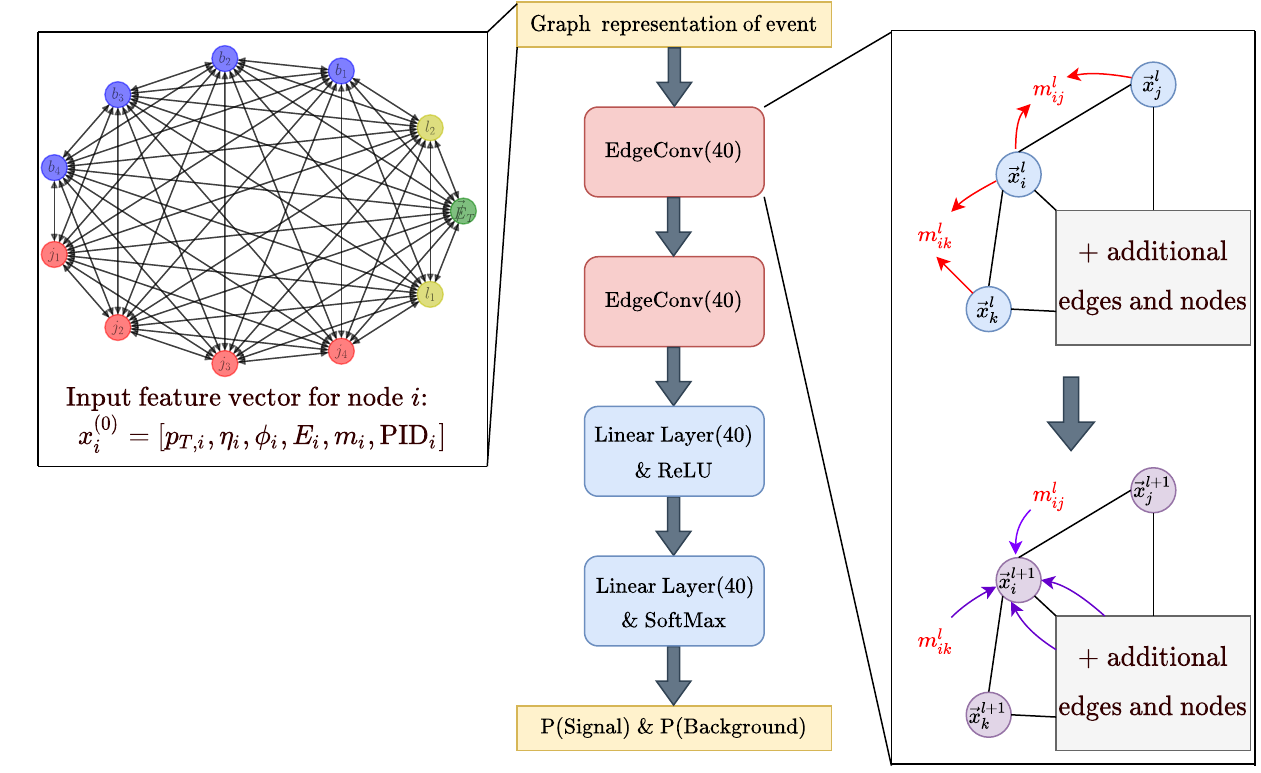}
	\caption{Example of a fully connected event graph (figure taken from ref.~\cite{Anisha:2023xmh}).}
	\label{fig:fourtop} 
\end{figure}

Graphs can have varying numbers of nodes and edges for each event, unlike fixed-size inputs in typical CNNs. Therefore, GNNs are adaptable to work on different event topologies and can generalize well across various types of particle collisions. In an early study, \cite{Abdughani:2018wrw} collision events were represented as event graphs, and the message-passing neural network (MPNN) was used to search for the stop pair production at the LHC. Employing GNN for particle-flow (PF) reconstruction in a high-pileup environment has been developed for a general-purpose multilayered particle detector \cite{Pata:2021oez}. In ref.~\cite{Anisha:2022ctm}, GNN with a fully connected graph is used to combat comprehensive backgrounds arising from other SM processes, which improves the sensitivity of di-Higgs analysis in the vector-boson fusion production channel. Similar architectures are used to improve the four top measurements for SM \cite{Anisha:2023xmh} and constrain the BSM parameter in the context of 2HDM. The fully connected graph and architecture used in the four top analysis are illustrated in figure \ref{fig:fourtop}. Self-attention to extract features from jet substructure and cross attention mechanism to combine these features for event level classification have also been studied~\cite{Hammad:2023sbd} for BSM searches at LHC showing improved performance compared to simple concatenation of features. Accurate measurement of the trilinear and quartic Higgs self-couplings is essential to understanding the shape of the Higgs potential and the electroweak phase transition. While traditional searches at the LHC using Higgs pair production provide limited constraints, recent studies using Graph Neural Networks (GNNs) have shown promise in improving these bounds \cite{Stylianou:2023xit}. Topological graph algorithms were also developed to reconstruct intermediate particles and address combinatorial challenges in full hadronic decays $t\bar{t}$ \cite{Ehrke:2023cpn}. Recently, hypergraph representations \cite{Birch-Sykes:2024gij} have been employed to demonstrate the reconstruction of parent particles for full hadronic decays of $t\bar{t}$. Events represented by weighted nodes with edges depicting the distance between events in kinematic space, were utilised~\cite{Mullin:2019mmh} to separate the stop quark pairs from the top pair in the kinematic space. In refs.~\cite{Pfeffer:2024tjl,Builtjes:2022usj}, $t\bar{t}$X process is scrutinized with Transformer and graph-based event classification. Using the point cloud representation for Higgs boson decays in tau leptons significantly improves compared to classical analysis techniques \cite{Onyisi:2022hdh}.
 
\paragraph{Anomaly Detection:} 
Even though there are many theoretically motivated extensions of the Standard Model, the search for new physics can be regarded as an inherently model-independent endeavour, as nature need not fit into the constraints of our imagination. This motivates developing algorithms that do not rely on a specific BSM model but try to find deviations from the known background SM processes. General techniques exist in ML literature which are appropriate for model-independent searches called anomaly detection, wherein an algorithm focuses on learning the features of a well-understood class (the background in our case) and identifying anomalous data samples that do not have the same features. 

A common way to achieve this unsupervised learning is via autoencoders. An autoencoder is an ANN used in unsupervised learning that aims to encode input data into a reduced-dimensional representation (called the latent representation) and then decode it back to its original form. The model learns by comparing the reconstructed output with the original input and tuning parameters to minimize the reconstruction error. Potential of autoencoder architectures have been demonstrated in simulating realistic and diverse aspects of high energy physics events. For instance, convolutional autoencoders were studied for detecting anomalous jet images in refs.~ \cite{Heimel:2018mkt,Farina:2018fyg}.

GNNs rely on understanding the graph’s topology and node connections to extract meaningful representations. Capturing this information accurately during the encoding and decoding phases to reconstruct the original graph structure is non-trivial. Particle Graph Autoencoder \cite{Kasieczka:2021xcg,Tsan:2021brw} used an edge-convolution encoder block to map the node features to a two-dimensional latent node representation, which fed to the symmetrical decoder block to reconstruct the node features. For anomaly detection of events or jets, we are interested in the global features of the graph. Although a graph readout extracts this information in classification, the operation itself is destructive, and one cannot recover the graph structure in an autoencoder based approach where the decoder needs to have the graph structure intact for its reconstruction. To address these challenges in graph autoencoders, ref.~\cite{Atkinson:2021nlt} devised edge reconstruction networks, which were used to reconstruct weighted edge features, thereby learning the graph structure without undergoing a destructive graph readout operation. The Deep Sets autoencoder \cite{Ostdiek:2021bem}, on the other hand, relied solely on the encoded global latent space without using any decoder. Graph autoencoders with physics-motivated inductive biases like Lorentz group equivariance \cite{Hao:2022zns} and IRC-safety  \cite{Atkinson:2022uzb} have also been explored for anomaly detection.


\section{Summary} 
\label{sec:summ} 

In this review, we have laid down the rationale for designing automatic feature extractors from high-dimensional data in the context of high-energy physics. The main theme of such strategies is the balance between expressivity and generalisation power. While expressivity deals with the theoretical power of non-parametric models to capture a large  set of functions accurately, generalisation power trades some of this expressivity for the practical gain of finding a good function for specific purposes based on prior knowledge of the use case. 

The wealth of fundamental understanding of particle physics motivates their use in designing automatic feature extractors for various applications at the Large Hadron Collider. One recurrent theme from prior physics intuition is the suitability of the set-based point cloud and graph representation of collision events. Among different deep learning frameworks, graph neural networks possess a unique ability to learn from relational data. They are also adaptable to versatile data structures, capturing the essence of graphs characterised by node features and their relationships (features related to connected edges) along with graph features, if any. At the heart of a GNN lies a clever mechanism called message passing. Nodes exchange information with their neighbours, iteratively aggregating and updating their representations. Such operations allow a node to understand its self-attributes in light of the context provided by its connected neighbours, capturing complex relationships within the graph.

It is not an understatement to state that mainstream scientific inquiry is in a data-driven era, propelled by the development of powerful and data-hungry deep-learning algorithms. However, from the perspective of fundamental fields like particle physics, the main objective is to update our understanding of the universe based on empirical evidence. Focusing on their performance alone would be detrimental in the long term, and the common consensus in the community is to understand these algorithms in sufficient detail. Removing this obscurity is not just of academic interest but rather a practical requirement arising from the infamous mathematical complexity of such algorithms. The (potential) discovery of new physics at the Large Hadron Collider is an extreme statistical condition, and one needs to understand the algorithm's behaviour in such extreme conditions concretely. 

\subsection*{Acknowledgements}
A.B. is supported by the U.S. Department
of Energy under grant number DE-SC 0016013.  V. S. N. is supported by the STFC under grant ST/P001246/1.  The computational works are performed using the Param Vikram-1000
High-Performance Computing Cluster and the TDP project resources of
the Physical Research Laboratory (PRL).
Authors would like to thank for the warm hospitality and support
received at the school of Deep Machine Learning for particle and
astroparticle physics (ML4HEP) at ICTS, Bengaluru (2023) and
IOP, Bhubaneswar (2024).
\vspace{.5cm}

\noindent 
\textbf{Data Availability Statement:} No data associated in the manuscript.
\bibliographystyle{JHEP}
\bibliography{ref}
	
\end{document}